# Stability of persistent currents in superfluid fermionic rings

Klejdja Xhani [1,*] Andrea Barresi [2,†] Marek Tylutki [2,‡] Gabriel Wlazłowski [2,3,§] and Piotr Magierski [2,3,‖]

[1]*Dipartimento di Fisica e Astronomia, Università di Bologna, Via Irnerio 46, 40127 Bologna, Italy*

[2]*Faculty of Physics, Warsaw University of Technology, Ulica Koszykowa 75, 00-662 Warsaw, Poland*

[3]*Department of Physics, University of Washington, Seattle, Washington 98195–1560, USA*

We investigate the stability of persistent currents in superfluid fermionic gases confined to a ring geometry. Our study, conducted at zero temperature using time-dependent density functional theory, cover interaction regimes from strong (unitary Fermi gas) to weak (Bardeen-Cooper-Schrieffer regime) couplings. Stability is tested against the presence of an external defect within the ring. The dissipation mechanism associated with vortex generation is present in all interaction regimes. Vortex emission is accompanied by Cooper pair breaking, which occurs even beyond the vortex core in the weakly interacting regime. The pair-breaking mechanism prevents the imprinting of a persistent current with a winding number above a threshold, which decreases as the system approach the BCS regime. Our study reveals the existence of two types of critical winding numbers above which currents cease to be persistent in Fermi superfluids: one related to the proliferation of quantum vortices and the other to the onset of the pair-breaking mechanism.

## I. INTRODUCTION

Persistent currents, the continuous flow of particles without dissipation in ring-shaped or similarly closed-loop traps, are fundamental phenomena that arise in quantum systems such as superfluids [1–3] and superconductors [4]. In the field of ultracold atomic physics, their study stands out as a pivotal area of research [5], promising deeper insight into quantum coherence and offering potential applications in quantum technologies [6,7]. The requirement for a single-valued wave function means the phase change around the loop must be an integer multiple of $2\pi$, defining a series of metastable states corresponding to the (integer) winding numbers $w$, separated by energy barriers [1]. The loop within which the persistent current flows corresponds to the case of a constant winding number. However, if dissipative effects are present, transitions to lower circulation state ma occur, like $w \rightarrow w - 1$. They can be induced by quantum/thermal phase slippage or by increasing the superfluid velocity beyond a certain critical value [8–14]. The dissipative effects vary depending on the nature of the superfluid. In condensed ultracold atoms of bosonic type, persistent current dissipation often arises from vortex proliferation within the superfluid, causing phase slippage and decay of the circulation [8,9,15,16]. The stability of these currents has been extensively studied, both with and without external defects [8,17–21]. In fermionic superfluids, the pair-breaking mechanism can significantly contribute to supercurrent dissipation [22], as seen in Josephson junction experiments [23,24]. Persistent currents in fermionic systems have been investigated theoretically [25–27] and were only recently observed experimentally [9,28]. In an experiment with ultracold atoms of $^6$Li [9], the effect of interaction strength on the critical winding number in the presence of external defects was examined. These studies attributed the dissipation of the current to vortex emission, finding that the gas tuned to a strongly interacting regime, referred to as a unitary Fermi gas (UFG), is particularly robust against the decay of persistent currents, making it a strong candidate for applications in quantum sensing.

Despite these advancements, several questions about the dissipation mechanisms in ultracold Fermi systems of the persistent currents remain unanswered. Quantum vortices can be imaged in the experiments, and it is relatively easy to correlate their emergence with the current decay, which is not the case for the second possible mechanism of Cooper pair breaking. This leads to questions about the importance of these mechanisms as a function of the interaction strength, which is hard to establish directly from the experimental data. A crucial aspect is identifying the pair-breaking critical winding number ($w_{\mathrm{pb}}$), beyond which Cooper pair breaking becomes energetically favorable. Understanding the impact of unpaired particles or a normal component on flow energy dissipation is crucial in the study of the stability of persistent currents. Furthermore, external factors such as defects or impurities can enhance Cooper pair breaking by acting as seed sites for pair-breaking events and promoting vortex emission. Comprehending the interplay between these factors is essential for predicting and controlling persistent currents in fermionic systems.

*Contact author: klejdja.xhani@unibo.it

†Contact author: andrea.barresi.dokt@pw.edu.pl

‡Contact author: marek.tylutki@pw.edu.pl

§Contact author: gabriel.wlazlowski@pw.edu.pl

‖Contact author: piotr.magierski@pw.edu.pl

This theoretical study explores the time-dependent dynamics of fermionic persistent currents in a two-dimensional ring geometry, similar to the experimental setup of Ref. [9], across various interaction strengths from strong to weak attractive superfluid regimes. We focus on clarifying how pair-breaking mechanisms, particularly relevant in the BCS regime [24], affect persistent current states and flow energy. We investigate whether the presence of a normal component or unpaired particles can induce transitions between metastable states and alter the energy landscape. Through a comprehensive analysis of both ground states and time-dependent evolutions of fermionic superfluids with imprinted currents, we identify two critical winding number values: the pair-breaking threshold ($w_{\rm pb}$) and the vortex emission threshold ($w_c$).

## II. THEORETICAL MODEL

We study the persistent current dynamics at different interaction strengths, from the strongly interacting unitary limit, characterized by a diverging $s$-wave interaction strength, to a weakly attractive interacting superfluid known as the Bardeen-Cooper-Schrieffer (BCS) regime. The superfluid dynamics is studied using density functional theory (DFT) techniques for superfluid Fermi gases [29–31]. In our computation, we use energy density functional known as the extended superfluid local density approximation (SLDAE) [32]. The static variant of this local DFT theory is formally equivalent to the mean-field Bogoliubov–de Gennes equations where the quasiparticles wavefunctions satisfy the equation

$$\mathcal{H}[\rho(\vec{r}), \nu(\vec{r})]\begin{pmatrix} u_n(\vec{r}) \\ v_n(\vec{r}) \end{pmatrix} = E_n \begin{pmatrix} u_n(\vec{r}) \\ v_n(\vec{r}) \end{pmatrix} \tag{1}$$

with $(u_n(\vec{r}), v_n(\vec{r}))^T$ being the Bogoliubov amplitudes and $E_n$ the quasiparticles states energy. The Hamiltonian $\mathcal{H}[\rho, \nu]$ is a function of the normal $\rho$ and anomalous $\nu$ densities, defined as

$$\rho(\vec{r}) = 2\sum_{E_n>0} |v_n(\vec{r})|^2, \tag{2a}$$

$$\nu(\vec{r}) = \sum_{E_n>0} u_n(\vec{r})v_n^*(\vec{r}), \tag{2b}$$

with the sum evaluated up to a cut-off energy $E_c$ in order to regularize the ultraviolet divergence [29,33]. We use the metric system, where $m = \hbar = 1$. The Hamiltonian has a generic form

$$\mathcal{H} = \begin{pmatrix} h(\vec{r}) - \mu & \Delta(\vec{r}) \\ \Delta^*(\vec{r}) & -h^*(\vec{r}) + \mu \end{pmatrix}, \tag{3}$$

where single particle Hamiltonian and pairing potentials are computed as appropriate functional derivatives of the energy functional $\mathcal{E}$, namely,

$$h = -\frac{1}{2}\nabla^2 + \frac{\delta\mathcal{E}}{\delta\rho} + V_{\rm ext}(\vec{r}), \tag{4}$$

$$\Delta = -\frac{\delta\mathcal{E}}{\delta\nu^*}. \tag{5}$$

The chemical potential is denoted as $\mu$. The mean-field $U = \delta\mathcal{E}/\delta\rho$ and pairing $\Delta$ potentials are functions of the dimensionless coupling constant $\lambda = a_s k_F$, where $a_s$ is the $s$-wave scattering length and $k_F = (3\pi^2\rho)^{1/3}$ is the Fermi wave vector. The SLDAE description ensures correct reproduction of the equation of state $E(\lambda)$ and strength of the paring gap $|\Delta(\lambda)|$. For the explicit form of the energy functional $\mathcal{E}$ and associated mean-fields, see Ref. [32]. The static approach allows us to determine properties of a state representing persistent current in a ring of a given winding number.

The time-dependent framework is obtained by allowing Bogoliubov amplitudes to be time dependent, i.e., $u_n(\vec{r}, t)$ and $v_n(\vec{r}, t)$, which also makes the densities time-dependent quantities, and by replacing $E_n \to i\frac{\partial}{\partial t}$ in Eq. (1). The model has been successfully applied to describe different physical phenomena such as the Josephson effect and dissipative mechanisms in a Josephson junction [24], dissipative vortices dynamics [34–39], Higgs modes [40], properties of spin-imbalanced systems [30,41–45], and even quantum turbulence [46–48]. The time-dependent framework is used to investigate the temporal behavior of the imprinted current, in particular its decay mechanism.

In our setup, the superfluid is trapped in a two-dimensional ring potential; see for visualization Fig. 1(c). The density is constant along the ring and goes to zero within a few coherence lengths at the edges. We impose translational symmetry along the $z$ direction, so the corresponding solutions are plane waves. We then imprint a phase $\phi = w_0 \arctan(x, y)$ in the order parameter, so that $\Delta(x, y) = |\Delta(x, y)|e^{i\phi}$. In this way, a persistent current state corresponding to the winding number $w_0$ is imprinted. Next, we study the stability of the persistent currents in the presence of a localized Gaussian defect by means of a time-dependent framework; see, for example, the inset of Fig. 3(a.i) below. The explicit form of the external potential is given in Appendix.

Our studies focus on three interaction regimes: unitary Fermi gas regime $\lambda^{-1} \simeq 0$ (labeled hereafter as UFG), experimentally accessible regime $\lambda^{-1} = -0.4$ (BCS limit) and $\lambda^{-1} \simeq -1$, which is presently inaccessible for experiments owing to very low value of critical temperature (deep BCS or dBCS). The main quantity that we use to characterize the stability of the current is the flow energy

$$E_{\rm flow}(t) = \int \frac{\vec{j}^2(\vec{r}, t)}{2\rho(\vec{r}, t)}\, d^3\vec{r}, \tag{6}$$

where

$$\vec{j}(\vec{r}) = 2\sum_{E_n>0} {\rm Im}(v_n \vec{\nabla} v_n^*) = \rho(\vec{r})\vec{v}(\vec{r}) \tag{7}$$

is the density current, which we also use to define velocity field $\vec{v}$. Note that, in general, $\vec{v}$ is not the same as superfluid velocity $\vec{v}_s = \frac{\hbar}{2m}\nabla\phi$, since the framework accounts also for effects (like pair-breaking) that effectively lead to the emergence of a normal component. In the simulations, the total energy is conserved. However, the flow energy can change in time, indicating current decay if $E_{\rm flow}$ is decreasing. The next quantity that we extract throughout our studies is the condensation energy, defined as

$$E_{\rm cond}(t) = \frac{3}{8}\int \frac{|\Delta(\vec{r}, t)|^2}{\varepsilon_F(\vec{r}, t)}\rho(\vec{r}, t)\, d^3\vec{r} \tag{8}$$

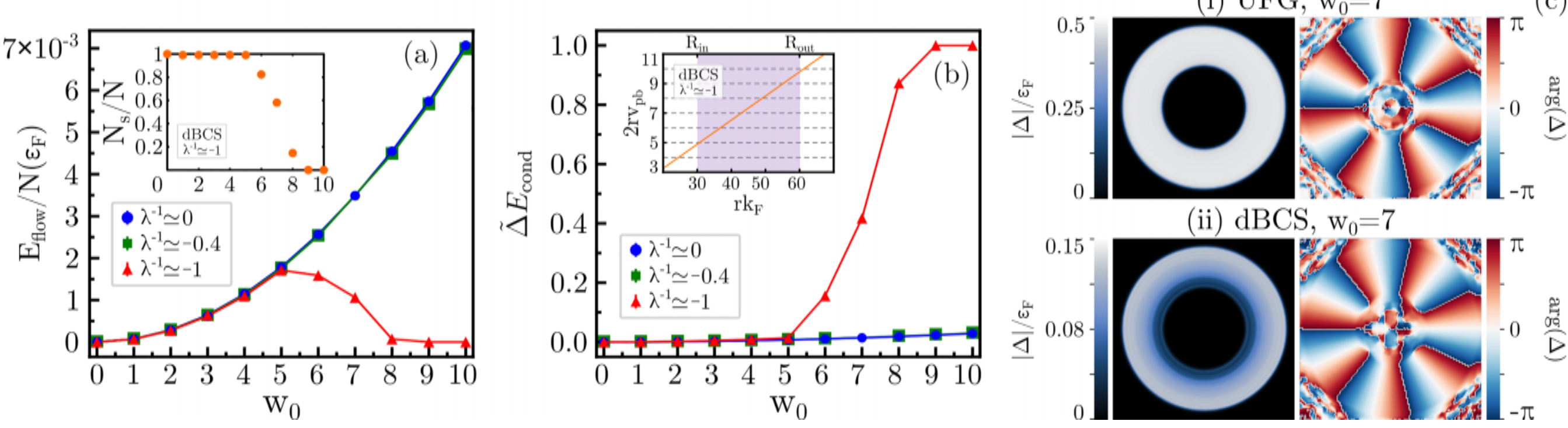

FIG. 1. Static calculations. The flow energy per particle number (a) and the relative condensation energy change $\tilde{\Delta}E_{\rm cond}$ (b) as a function of the imprinted winding number $w_0$ from the UFG limit ($\lambda^{-1} \simeq 0$, blue dots) to deep-BCS ($\lambda^{-1} \simeq -1$, red triangles), with BCS ($\lambda^{-1} \simeq -0.4$, green squares) being an intermediate regime. The inset in (a) shows the superfluid fraction as a function of winding number $w_0$ for the deep-BCS regime. The flow energy increases with $w_0$ as $E_{\rm flow} \propto w_0^2$ in UFG and BCS limit for all $w_0$ and for values of $w_0 \leqslant w_{\rm pb} = 5$ in dBCS regime as in the latter case the pair-breaking mechanism comes into play even in the absence of the defect. This occurs when the imprinted velocity field $w_0/2r$ starts to exceed pair-breaking velocity $v_{\rm pb}$. In the inset of (b), we show the corresponding value of $w_0 = 2rv_{\rm pb}$ at which this happens as a function of distance from the ring center for the dBCS case. The pair-breaking mechanism becomes more effective when $w_0$ is increased, affecting first the area close to the inner radius $R_{\rm in}$ and eventually leading to a loss of superfluidity in the whole volume. The panel (c) shows the absolute value of the order parameter (left column) and its phase (right column) in the $xy$ plane in UFG (i) and deep-BCS (ii) regimes. The two-dimensional phase plots correspond to the imprinted winding number $w_0 = 7$.

with $\varepsilon_F(\vec{r})$ being the local value of the Fermi energy [24]. It measures the amount of energy contained in the condensate of Cooper pairs and its decrease in time is a measure of the importance of pair-breaking mechanisms.

## III. STATIC CALCULATIONS

For each interaction regime, we perform a systematic study of the dependence of the flow and the condensation energy with respect to the initial winding number $w_0$, as shown in Figs. 1(a) and 1(b), respectively. Assuming that the system is a pure superfluid, $\rho = \rho_s$ (no broken Cooper pairs), and the velocity field is equivalent to the superfluid velocity, $v = v_s$, the flow energy reads

$$E_{\rm flow} = 2\pi L_z \int_{R_{\rm in}}^{R_{\rm out}} \frac{\rho_s(r) v_s^2(r)}{2}\, r dr = \frac{\pi L_z \rho_s w_0^2}{4} \ln\left(R_{\rm out}/R_{\rm in}\right), \tag{9}$$

where $R_{\rm in} = 30k_F^{-1}$, $R_{\rm out} = 60k_F^{-1}$ are the inner and outer radius of the ring within which the density is constant, and $L_z$ is the size of the system along the $z$ direction. The superfluid velocity induced by the imprinted phase profile is $v_s(r) = w_0\hbar/(2mr) = w_0/2r$ (in our units). Notably, as long as the system remains purely superfluid, the flow energy remains independent of the interaction regime and scales as a square of the winding number $w_0$. This can be seen for UFG and BCS in the entire range of considered winding number values, while for dBCS simulations only for $w_0 \leqslant 5$, see Fig. 1(a). The departure from $\sim w_0^2$ scaling above $w_0 > 5$ is a clear signature of the breakdown of this assumption. We have verified which fraction of the system responds to the phase imprinting. The imprinting procedure should induce a current only in the superfluid component $j(\vec{r}) = \rho_s(\vec{r})v_s(\vec{r})$, i.e., we assume that the normal component is not affected by it and remains at rest. Under this assumption, we can extract the number of atoms belonging to the superfluid component

$$N_s = \int \rho_s(\vec{r}) d^3\vec{r} = \int \frac{2r}{w_0} j(\vec{r}) d^3\vec{r}, \tag{10}$$

and compare it to the total number of atoms $N = \int \rho(\vec{r}) d^3\vec{r}$, whose ratio is shown in the inset of Fig. 1(a). Thus, the decrease of the $E_{\rm flow}$ for $w_0 > w_{\rm pb}$ is caused by the decreasing superfluid fraction. In particular, the former drops to zero for $w_0 \geqslant 9$ indicating that the imprinted phase no longer induces a flow, which is consistently reflected by vanishing superfluid fraction—an unambiguous demonstration that the system is in the normal state.

Further insight is provided by an analysis of the relative change in the condensation energy with respect to the state without imprinted current ($w = 0$), $\tilde{\Delta}E_{\rm cond}(w_0) = \frac{|E_{\rm cond}(w_0) - E_{\rm cond}(w_0=0)|}{E_{\rm cond}(w_0=0)}$, see Fig. 1(b). A drop in the condensation energy signals the presence of a pair-breaking mechanism. In the UFG and BCS limits $\tilde{\Delta}E_{\rm cond}(w_0)$ increases slightly with $w_0$, remaining below 5% even at the highest explored value of $w_0 = 10$, which means that the pair-breaking mechanism is negligibly small. However, in the dBCS regime, $\tilde{\Delta}E_{\rm cond}$ rises rapidly above $w_0 = 5$, allowing us to introduce the critical winding number $w_{\rm pb}$, above which the broken pairs become essential contributors to the decay of a persistent current.

Next, we estimate the pair-breaking threshold $v_{\rm pb}$, defined as [24,49]

$$v_{\rm pb} = \sqrt{\sqrt{\mu^2 + \Delta^2} - \mu}, \tag{11}$$

and compare it to the expected value of velocity owing to the phase imprinting $v_s(\boldsymbol{r})$. Note that $v_{\rm pb}$ is the property of the system, while $v_s(r) = \frac{w_0}{2r}$ depends on the distance from the system's center. In general, for a fixed $w_0$, we may have a situation where, in some parts of the system, the $v_s$ exceeds pair-breaking velocity, while in others, it does not. In the inset of Fig. 1(b), we show the quantity $w_0 = 2rv_{\text{pb}}$ for the deep-BCS case, from which we determine the minimal value for the winding number needed to surpass the pair-breaking velocity at a given distance $r$ from the center. We immediately find that the superflow starts to break Cooper pairs first in the vicinity of the inner ring, and in the dBCS case, it happens for winding number $w_{\text{pb}} \approx 2R_{\text{in}}v_{\text{pb}} \approx 5$. This value matches perfectly with changing of trends for $E_{\text{flow}}$ and $\tilde{\Delta}E_{\text{cond}}$ in Fig. 1. Thus, we identify $w_{\text{pb}} = 5$ as the *pair-breaking winding number* for the dBCS regime. As we increase $w_0$, the local velocity $v$ exceeds $v_{\text{pb}}$ even at a greater distance from the inner edge of the ring, and eventually, the superfluidity is completely lost. For example, in the dBCS case, $v_s(r)$ exceeds pair-breaking velocity $v_{\text{pb}}$ for all $r \in (R_{\text{in}}, R_{\text{out}})$ if $w_0 \gtrsim 9$, being again in agreement with our simulations. Figure 1(c) shows the spatial distribution of the order parameter for a fixed winding number $w_0 = 7$ for UFG and dBCS cases. In the former, the induced velocity field does not exceed $v_{\text{pb}}$ in any portion of the ring, and the order parameter remains unaffected, while in the latter one, the flow induces broken pairs near the inner edge. It is manifested by the suppression of the magnitude of the order parameter in that part of the ring. In summary, the value of the $w_{\text{pb}}$ depends both on the interaction regime, since $v_{\text{pb}} = v_{\text{pb}}(\lambda)$, and on the geometry of the system.

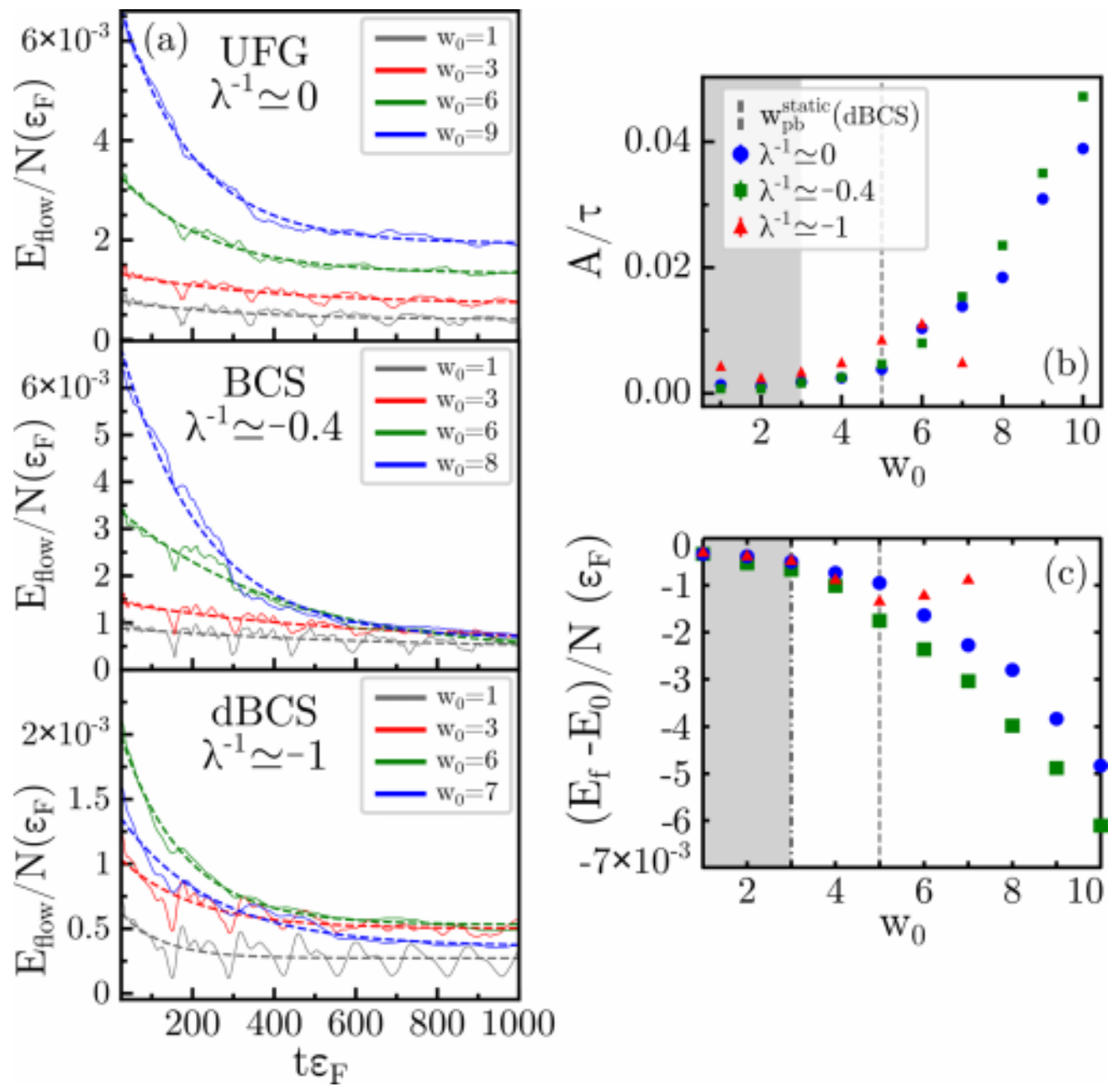

FIG. 2. Time-dependent calculations. (a) Temporal evolution of the flow energy for selected values of winding number $w_0$ and for different interaction strengths $\lambda^{-1}$ from UFG to dBCS limit, where the time is expressed in units of $\varepsilon_F^{-1}$. The dashed lines are the exponential fit $F(t) = Ae^{-t/\tau} + B$. [(b),(c)] The dependence of the decay rate $A/\tau$ and the energy loss $E_f - E_0$ per particle as a function of winding number $w_0$. The shaded area indicates the values of the winding number for which only sound is emitted owing to the switch on the defect. The dashed line indicates the onset of pair-breaking phenomena in the deep-BCS regime extracted from the static considerations.

## IV. DYNAMIC CALCULATIONS

The static considerations demonstrated that the persistent currents could be affected significantly by Cooper pair breaking when a sufficiently large initial flow is imprinted. However, they do not provide information about the stability of the currents once they are imprinted. To investigate this, we perform their time-dependent evolution. We study the stability in the presence of the defect, as was done in the experiment [9]. The defect's height is switched on until it reaches the final value of $V_0 \simeq 2\mu$ at $t_0 = 25\varepsilon_F^{-1}$, having a width at $1/e^2$ of $\sigma = 10k_F^{-1}$. The defect's parameters are kept fixed in all explored interaction regimes. The stability is judged based on the measurement of the temporal profile of the flow energy $E_{\text{flow}}(t)$ after the defect is switched on, see Fig. 2(a). These data are fitted via an exponential function given as

$$E_{\text{flow}}(t) = Ae^{-t/\tau} + B, \tag{12}$$

where $A$, $B$, and $\tau$ are fitting parameters, with $\tau$ being the decay time. By taking the time derivative of Eq. (12), we note that $A/\tau$ gives the decay rate of the flow energy, Fig. 2(b). Next, we define a second quantity named the flow energy loss as $E_f - E_0$ with $E_f = E_{\text{flow}}(t \to \infty)$ and $E_0 = E_{\text{flow}}(t_0)$ denoting the final and initial flow energies respectively, Fig. 2(c). At low $w_0$ values, the flow energy tends to remain relatively stable over time, with fluctuations caused by the defect activation procedure. However, as $w_0$ increases, the decay rate $A/\tau$ becomes larger, as illustrated in Fig. 2(b). Consequently, the amount of the energy loss also increases with $w_0$, as depicted in Fig. 2(c). Interestingly, while we observe a similar trend in the dBCS case for $w_0$ values up to $w_0 = 5$, flow energy dissipation decreases for even higher winding numbers. It reflects the fact that the initial reservoir of the flow energy $E_{\text{flow}}(t = 0)$ begins to shrink for $w_0 > 5$, owing to the induction of the normal component as shown in Fig. 1(a), and thus the amount of energy to be dissipated $(E_f - E_0)$ shrinks as well. Furthermore, we find that dissipation in the BCS regime surpasses that in the UFG limit for equivalent $w_0$ values. This observation suggests the presence of mechanisms that amplify dissipation, specifically in the BCS regime. Consequently, we investigate the origins of dissipation in all three regimes and examine how it varies with initial winding number values.

Our analysis reveals two primary phenomena contributing to the dissipation of superflow energy: *the generation of vortices* and/or *the pair-breaking phenomena*. The former is revealed by the decay of the winding number over time: whenever a vortex is generated, it causes a phase slippage and removes one unit of the circulation. Thus, the instantaneous winding number (measured as the number of jumps by the phase by $2\pi$ as we move along the inner edge of the ring) changes in discrete steps, see Fig. 3(a). The latter contribution is instead discerned from the temporal behavior of the condensation energy, as illustrated in Fig. 3(b). Both phenomena lead to temporal changes in flow energy. An intriguing discovery is that the critical winding number $w_c$, denoted as the last value of $w_0$ for which $w(t) = w_0$, appears to be independent of the interaction regime, consistently standing at the value of $w_c = 2$. For $w_0 \geqslant 3$, we observe the generation of quantum vortices. They are nucleated near the defect and may eventually propagate into the bulk, as shown in the insets of Figs. 3(a.i) and 3(a.ii), with 2D profiles of $|\Delta|$ at selected moments and in Ref. [50]. Their number is related to $(w(t) - w_0)$ value. Figure 3(a.iii) shows the final winding number, extracted by an exponential fit of the $w(t)$, as a function of $w_0$. As $w_0$ increases, the difference $(w_0 - w_f)$ also increases, indicating the total amount of deposited vortices. For pure superfluid ($\rho = \rho_s$), the flow energy is expected to be quadratically related to the winding number, so its decay in time clearly signals the energy loss. There is a striking feature of Fig. 3(a.iii) that requires clarification. No qualitative differences appear for $(w_0 - w_f)$ when comparing different interaction regimes. However, at the quantitative level, the plot suggests that more vortices are emitted and, therefore, expected enhanced dissipation in the UFG regime as compared to the BCS regime for $w_0 > 6$ (green square points are systematically above blue dots). It is in tension with data presented in Fig. 2(c), which points to the opposite.

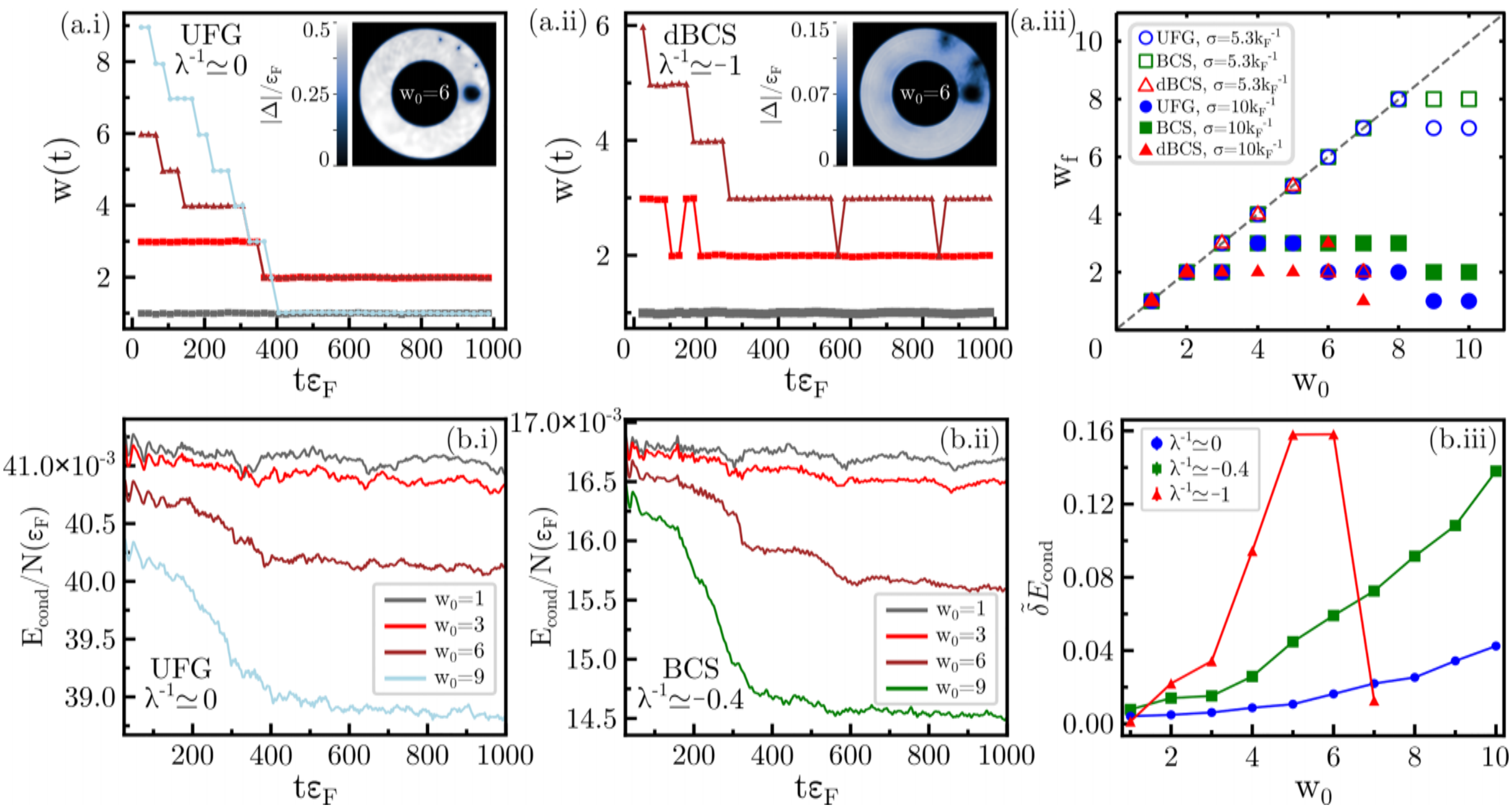

FIG. 3. The winding number, extracted at the ring's inner edge, as a function of time in UFG (a.i) and the dBCS regime (a.ii). The initial time $t_0$ corresponds to the moment at which the height of the potential simulating the defect reaches its final value $V_0/\mu \simeq 2$. Insets: color maps of $|\Delta(x,y)|/\varepsilon_F$ for $w_0 = 6$ at times when several vortices propagate into the bulk in UFG and dBCS limit for $t\varepsilon_F \simeq 430$ and $t\varepsilon_F = 320$, respectively. Subfigure (a.iii) shows the final winding number $w_f$ as a function of the imprinted winding number $w_0$. Filled symbols indicate data with larger defect $\sigma = 10k_F^{-1}$ and $V_0 = 2\mu$, whereas empty symbols represent data with smaller defect $\sigma = 5.33k_F^{-1}$ and $V_0 = 0.8\mu$. The dashed line indicates the ideal case where no phase-slippage events take place ($w_f = w_0$). Condensation energy per particle as a function of time in UFG (b.i) and BCS regime (b.ii) for selected initial winding numbers. (b.iii) The difference in relative condensation energy change as a function of the initial winding number for all three regimes (UFG, BCS, and dBCS).

To gain further insight into this puzzling problem, we investigate the temporal behavior of our second observable: the condensation energy per particle $E_{\text{cond}}(t)/N$, shown in Figs. 3(b.i) and 3(b.ii) for UFG and BCS limit. Given the scale differences between its values in different interaction regimes, we compare the data by taking the normalized difference between final and initial values of $E_{\text{cond}}$,

$$\tilde{\delta}E_{\text{cond}} = \frac{E_{\text{cond}}(t_f) - E_{\text{cond}}(0)}{E_{\text{cond}}(0)}, \tag{13}$$

as illustrated in Fig. 3(b.iii) for all three regimes. The apparent difference in condensation energy loss is clearly visible across different interaction regimes. Effectively, the loss can be interpreted as the production of the normal component at the cost of the flow energy. In general, this mechanism can operate independently. For example, we may have a situation where the spatial distribution of the phase of the order parameter (and thus superfluid velocity $v_s$) does not further change over time, while part of the superfluid density is converted into normal density, which dissipates the flow energy via viscosity. However, if quantum vortices are involved, one cannot decouple their generation from the pair-breaking. It is because of the peculiar structure of vortices in fermionic superfluids: the disruption of Cooper pairs in vortex cores is induced by velocities of the superflow approaching the pair-breaking velocity [shown in Fig. 4(a)]. When sufficient number of vortices are created, the portion of the superfluid volume in which $|\Delta| \neq 0$ decreases, and $E_{\text{cond}}$ follows suit. The drop of the condensation energy, as seen in Fig. 3(b.i) for UFG, is the result of this process. The larger $E_{\text{cond}}$ decrease for the dBCS regime is caused by the larger vortex core size [inset of Fig. 3(a.ii)], but besides that, we also observe the pair-breaking process in action away from the defect and from the vortex cores, see Fig. 4(b). We find the systematic increase in the importance of the pair-breaking mechanics as we move from unitarity to BCS regimes, as demonstrated in Fig. 4(c).

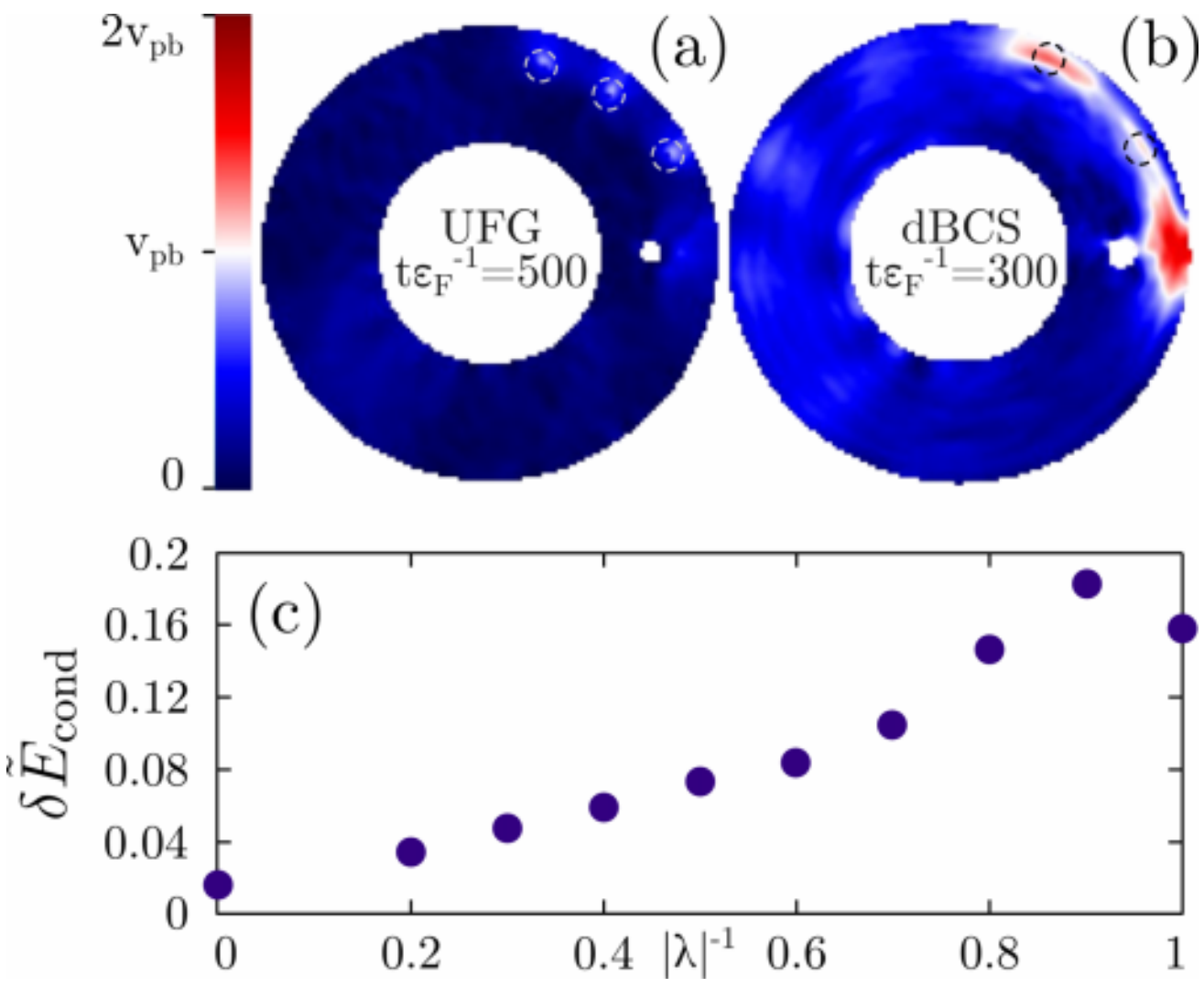

FIG. 4. (a) Two-dimensional velocity field values, computed according to Eq. (7), at unitarity ($\lambda^{-1} \simeq 0$) for $t\varepsilon_F \simeq 500$ and in dBCS regime ($\lambda^{-1} \simeq -1$) and (b) for $t\varepsilon_F \simeq 300$. Dashed circles indicate vortex cores. (c) Difference in condensation energy [as defined in Eq. (13)] as a function of $|\lambda|^{-1}$ for a chosen value of imprinted winding number ($w_0 = 6$).

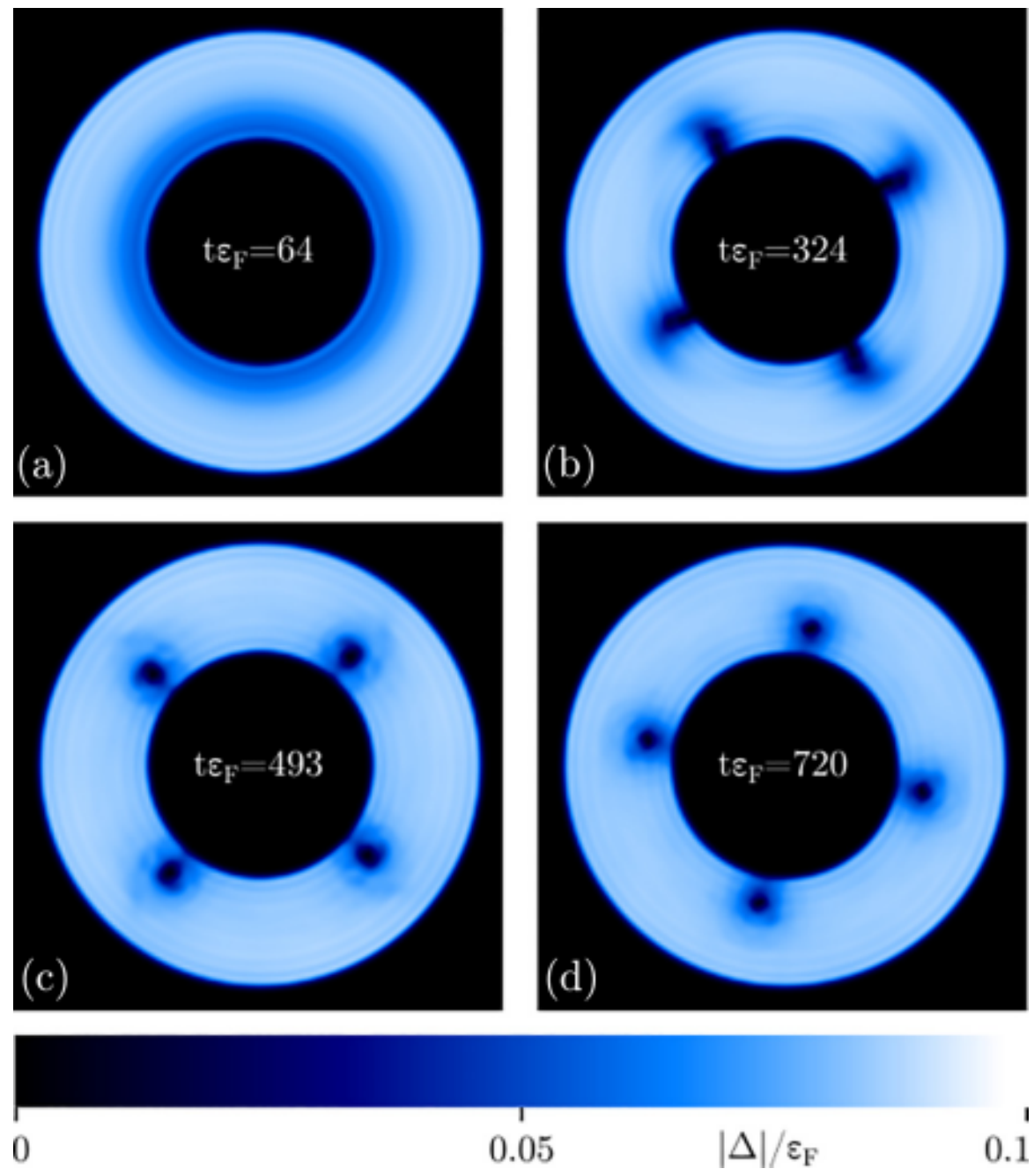

FIG. 5. Time evolution of $|\Delta|/\varepsilon_F$ for the deep-BCS regime ($\lambda^{-1} \simeq -1$) with initial $w_0 = 6$ and without the defect. (a) $t\varepsilon_F = 64$, the pairing gap is lower around the inner ring; (b) $t\varepsilon_F = 324$, vortex cores begin to emerge from the center; (c) $t\varepsilon_F = 493$, vortices begin moving along the direction of the flow; and (d) $t\varepsilon_F = 720$, rotation continues with the vortex cores maintaining the respective distance.

In the deep BCS regime, for the large initial values of $w_0$, there is already a normal component (broken pairs) in the initial state, which significantly impacts the dynamics from the beginning. When $w_0 \geqslant 6 > w_{\rm pb}$, the initial current state surpasses the pair-breaking critical velocity, leading to an initial partial loss of superfluidity. The latter promotes the emission of vortices during the dynamics, even far from the defect region. We observe vortex generation in this regime even in the absence of the defect. Figure 5 shows $|\Delta|/\varepsilon_F$ for different instants in time while following the vortices emission for $w_0 = 6$ and no defect. The presence of the normal component, concentrated mainly in the inner ring vicinity, together with the suppressed local value of $|\Delta|$, may by themselves act as a proliferator of the vortices. Furthermore, as the coherence length is given by $\xi = \frac{k_F}{\pi\Delta}$, lower local values of $|\Delta|$ lead to larger values of $\xi$. As a consequence, the defect has a negligible width $\sigma \ll \xi$, and thus, fewer vortices are emitted owing to the external defect.

If we further increase $w_0$ we almost kill the superfluidity ($\Delta \approx 0$), as seen, e.g., for $w_0 = 8$, and therefore $\tilde{\delta} E_{\rm cond}$ goes to zero. Winding numbers $w_0 = 9, 10$ generate a flow that significantly exceeds the pair-breaking velocity in the entire volume. The resulting initial states correspond to the system being in the normal state with no persistent currents; thus, we do not include these cases in the analysis.

Experimental investigations in Ref. [9], which featured defects considerably smaller than those in our study, found that $w_c$ depends on $\lambda$ with the UFG limit being more robust against dissipation via vortex emission than the BCS regime, contrasting with our findings. To clarify this discrepancy, we performed a new set of simulations with a smaller defect, specifically for $V_0 = 0.8\mu$ and $\sigma = 5.3k_F^{-1}$, closer to the experimental values. We find that the critical winding number varies across different interaction regimes: it is $w_c = 8$ at unitary and BCS regimes, while it drops to $w_c = 5$ in the dBCS regime, as shown in Fig. 3(a.iii) by open symbols. Thus, for the smaller defect we also find a dependence of $w_c$ with respect to the interaction regime, which is in qualitative agreement with the experimental findings [9]. The dependence of $w_c$ on the interaction regime is not universal and depends on the defect's parameters. In fact, consistently with predictions for the BEC regime [8], as the defect size is decreased, the critical winding number $w_c$ increases. As we increase $w_0$, if it exceeds $w_{\rm pb}$, the persistent current ceases to be persistent owing to emission of vortices throughout the inner edge of the ring, as shown in Fig. 5. Thus, the critical value $w_c$ cannot be higher than $w_{\rm pb}$, as found in the dBCS regime for our explored values of $w_0$. There are still discrepancies at the quantitative level between the experimental and theoretical results, like we find the value of $w_c = 5$ for $\lambda^{-1} \simeq -1$ while the experiment reports this value already for $\lambda^{-1} \simeq -0.4$. However we note that the experimental imprinting procedure was different from that we use in the simulations. In the experiment, the imprinting generates excitations, such as vortices and/or sound waves, which could favor the emission of vortices at lower $w_0$ values and thus lowering $w_c$. Another difference is related to the temperature. Our simulations were conducted at absolute zero, while the experiment operates for temperatures being approximately 40% of the critical temperature.

## V. CONCLUSIONS

Studying both static and dynamic properties of fermionic superfluids flowing in the ring reveals two distinct critical winding numbers. The first threshold is the pair-breaking winding number $w_{\text{pb}}$, where the local velocity surpasses the pair-breaking threshold, resulting in a decrease in the superfluid fraction. Notably, $w_{\text{pb}}$ is significantly lower in the deep-BCS regime compared to the unitarity regime, aligning with expectations. As long as the amount of the normal component is marginal, the flow energy increases with $w_0^2$. Towards the weakly attractive BCS limit, the breakup of Cooper-like pairs leads to a decline in both the superfluid fraction and the associated flow energy until superfluidity is lost. The second critical winding number ($w_c$) emerges in the presence of an external localized defect and represents the threshold for dynamic dissipation of the winding number through vortex generation. For smaller defects, it depends on the interaction regime with the UFG limit having the highest value, which is consistent with the experimental findings. Instead, it is almost independent of the interaction regime for larger defects.

We have explored the interplay between vortex emission and pair-breaking during superfluid dynamics. Both these phenomena directly impact the flow energy dissipation, with the dissipation rate increasing with $w_0$ and as we approach the BCS regime. Vortex emission notably influences the condensation energy owing to the higher density of unpaired particles within the vortex cores, but broken pairs can also be produced independently far away from vortex cores and the defect, which is particularly pronounced in the deep BCS regime. Conversely, the broken pairs formed by the phase imprinting method, which constitute a normal component, enhance the process of vortices emission during the dynamics even in the absence of an external defect as occurs in dBCS regime for $w_0 > w_{\text{pb}}$.

Our study shows that the measurement of the winding number's decay is insufficient to determine whether the current is persistent or decaying unambiguously. We identify situations where the flow energy decreases in certain time intervals, without a decay of $w(t)$. This is because observables based solely on the winding number, while sensitive to the generation of quantum vortices, are only weakly sensitive to the broken pairs.

From the perspective of atomtronics applications and superconductors, further exploration could involve quantitative analysis of the dissipative phenomena discussed here in the presence of multiple defects along the superflow. The study also points to the importance of searching for another probe to test the stability of the superfluid flows, such as those that quantify superfluid fraction during the time evolution.

## ACKNOWLEDGMENTS

We are grateful to G. Roati, G. Del Pace, D. Pęcak, and B. Tüzemen for their insightful discussions and ideas. The calculations were executed by means of the W-SLDA Toolkit [51]. M.T. was supported by a National Science Centre Grant No. UMO-2019/35/D/ST2/00201. A.B., G.W., and P.M. were supported by a National Science Centre Grant No. UMO-2022/45/B/ST2/00358. This work used computational resources `Tsubame3.0` supercomputer provided by Tokyo Institute of Technology through the HPCI System Research Project (Project ID: `hp230081`). We also gratefully acknowledge Polish high-performance computing infrastructure PLGrid (HPC Centers: ACK Cyfronet AGH) for providing computer facilities and support within Computational Grant No. PLG/2024/016930 (`plginhsf2`).

## DATA AVAILABILITY

Reproducibility packs for restoration of results presented in this paper are available via Zenodo repository [50].

## APPENDIX: SHAPE OF TRAPPING POTENTIAL AND DEFECT

The external potential is imprinted in an annular shape during the self-consistent procedure, defined by the following expression:

$$V_{\text{ext}}(\vec{r}) = \begin{cases} 2\varepsilon_F & r < R_{\text{in}} = 26k_F^{-1} \\ 2\varepsilon_F\big[1 - s\big(r - R_1^s, R_1^s - R_{\text{in}}\big)\big] & R_{\text{in}} \leqslant r \\ 0 & R_1^s \leqslant r < R_2^s = 56k_F^{-1} \\ 2\varepsilon_F s\big(r - R_2^s, R_{\text{out}} - R_2^s\big) & R_2^s \leqslant r < R_{\text{out}} = 60k_F^{-1} \\ 2\varepsilon_F & r \geqslant R_{\text{out}} \end{cases} \tag{A1}$$

where $s(a, A) = \frac{1}{2} + \frac{1}{2}\tanh\left[\tan[\frac{\pi}{2}(\frac{2a}{A} - 1)]\right]$ is a radial function that modifies the potential continuously between radii. This is done in order to prevent discontinuities, which would create numerical instabilities in the simulations. After the ground state has been determined, we dynamically raise an additional potential that models defect during the time-dependent part. The defect is placed at a position $(x_D, y_D) = (\frac{R_{\text{out}} - R_{\text{in}}}{2}, 0)$ and is taken to have a Gaussian shape $V(t) = f(t)V_0 e^{-2x_D^2/\sigma^2} e^{-2y_D^2/\sigma^2}$, where the defect widths we studied are: (a) $\sigma k_F = 10$ and height $V_0 = 2\mu$, and (b) $\sigma k_F = 5.3$ and height $V_0 = 0.8\mu$. The function $f(t)$ linearly rises from 0 to 1 in time interval $t \in (0, t_0 = 25\varepsilon_F^{-1})$, and for $t > t_0$ is kept to be 1.